\documentclass[pra,twocolumn]{revtex4-1}

\usepackage{amsmath}
\usepackage{amssymb}
\usepackage{graphicx}

\newcommand{\bra}[1]{\left\langle #1 \right|}
\newcommand{\ket}[1]{\left| #1 \right\rangle}

\begin{document}

\title{Classical capacity of Gaussian communication under a single noisy channel}

\author{Jaehak Lee} \email{jaehak.lee@qatar.tamu.edu}
\author{Se-Wan Ji}
\author{Jiyong Park}
\author{Hyunchul Nha} \email{hyunchul.nha@qatar.tamu.edu}
\affiliation{Department of Physics, Texas A \& M University at Qatar, P.O. Box 23874, Doha, Qatar}

\begin{abstract}
A long-standing problem on the classical capacity of bosonic Gaussian channels has recently been resolved by proving the minimum output entropy conjecture. It is also known that the ultimate capacity quantified by the Holevo bound can be achieved asymptotically by using an infinite number of channels. However, it is less understood to what extent the communication capacity can be reached if one uses a finite number of channels, which is a topic of practical importance.
In this paper, we study the capacity of Gaussian communication, i.e., employing Gaussian states and Gaussian measurements to encode and decode information under a single-channel use. We prove that the optimal capacity of single-channel Gaussian communication is achieved by one of two well-known protocols, i.e., coherent-state communication or squeezed-state communication, depending on the energy (number of photons) as well as the characteristics of the channel. Our result suggests that the coherent-state scheme known to achieve the ultimate information-theoretic capacity is not a practically optimal scheme for the case of using a finite number of channels. We find that overall the squeezed-state communication is optimal in a small-photon-number regime whereas the coherent-state communication performs better in a large-photon-number regime.
\end{abstract}

\maketitle

\section{\label{sec:introduction}Introduction}

The ultimate classical capacity of bosonic channels \cite{bib:RevModPhys.66.481, bib:PhysRevA.63.032312, bib:PhysRevLett.92.027902, bib:PhysRevA.70.032315, bib:NatPhoton.7.834} has been a long standing problem in quantum communication theory. The Holevo quantity \cite{bib:ProblInfTransm.9.177} provides an upper bound for the mutual information between the sender and the receiver, which thereby puts a limitation on the number of bits shared between communicators. In a communication protocol where a sender prepares a quantum state $\rho_i$ embedding a classical variable $x_i$ with probability $p_i$ and sends it to a receiver via a channel $\mathcal{E}$, the Holevo quantity is given by
\begin{equation} \label{eq:Holevo}
\chi(\mathcal{E}) = S\left(\sum_i p_i \mathcal{E}(\rho_i) \right) - \sum_i p_i S\left( \mathcal{E}(\rho_i) \right) ,
\end{equation}
where $S(\rho)$ represents von Neumann entropy of a quantum state $\rho$. 
For a given channel $\mathcal{E}$, the ultimate classical capacity is defined to be the regularized Holevo quantity optimized over the strategies with $\{p_i,\rho_i\}$---, the so-called Holevo bound, that is,
\begin{equation}
C(\mathcal{E}) = \lim_{m\to\infty} \frac{1}{m} \max_{\{p_i,\rho_i\}} \chi(\mathcal{E}^{\otimes m}),
\end{equation}
where $\mathcal{E}^{\otimes m}$ represents $m$ uses of the channel $\mathcal{E}$.

For a continuous-variable (CV) system, the above quantity can be an unlimitedly large number due to the infinite dimension; thus a practically relevant constraint is typically introduced, i.e., a finite energy of the system. Under this energy constraint, it is well known that the first term on the right-hand side of Eq. (\ref{eq:Holevo}) is maximized by a thermal state, $ \sum_i p_i \mathcal{E}(\rho_i) = \rho_\textrm{th} $, owing to the extremality of Gaussian states \cite{bib:PhysRevLett.96.080502}. Thus, in order to identify the ultimate capacity, it remains to minimize the second term, related to minimum output entropy conjecture \cite{bib:PhysRevA.70.032315}, which states that minimum entropy of the output state of a Gaussian bosonic channel is realized by a coherent state input. Recently, the conjecture was proven to be true, so the ultimate capacity of phase-insensitive Gaussian channels was completely obtained \cite{bib:NatPhoton.8.796,bib:NatCommun.5.3826}. Meanwhile, it was also shown that the capacity of phase-insensitive Gaussian channels is additive, that is, optimal encoding is separable \cite{bib:NatPhoton.8.796,bib:NatCommun.5.3826}.

On the other hand, it is a nontrivial problem to identify a decoding method, i.e., a measurement scheme at a receiver station, to achieve the Holevo quantity in Eq. (\ref{eq:Holevo}). In principle, the Holevo-Schumacher-Westmoreland (HSW) theorem \cite{bib:TransInf.44.269,bib:PhysRevA.56.131} showed that the Holevo bound can be achieved {\it asymptotically} by using an infinite number of channels for an arbitrary quantum channel. In the proof of the HSW theorem, they employed a collective measurement, the so-called square-root measurement, which requires highly nonlinear operations. In general, however, the Holevo bound may not be achieved with a single- or finite-channel communication only. Therefore, it is of crucial practical importance to study an optimal channel capacity under a finite-channel use \cite{bib:PhysRevA.58.146,bib:PhysRevA.61.032309,bib:PhysRevLett.106.240502} and identify the gap between this practically realizable capacity and the information-theoretic Holevo bound. Let us take an example of a loss channel. The capacity of the thermal-loss channel is given by \cite{bib:PhysRevLett.92.027902}
\begin{eqnarray} \label{eq:losscapacity}
C_\textrm{loss} & = & g \left[ \eta\bar{n} + (1-\eta)n_\textrm{th} \right] - g \left[ (1-\eta)n_\textrm{th} \right] , \\
g(x) & \equiv & (1+x)\log_2(1+x) - x\log_2 x , \nonumber
\end{eqnarray}
where $\eta$ represents the transmissivity of the channel, $\bar{n}$ the average photon number per channel use, $n_\textrm{th}$ the photon number of the environment, and $g(\bar{n})$ is the von Neumann entropy of a thermal state. (Throughout this paper, the logarithm is taken to the base 2 so that information is measured in number of bits.) For a perfect channel, i.e., $ \eta = 1 $, the capacity is achieved by a number-state communication \cite{bib:RevModPhys.66.481,bib:PhysRevLett.70.363}, where variable $x_i$ is encoded in number state $\ket{i}$ according to a thermal distribution and measured with a photon-number-resolving detector. If the channel is not ideal, however, the number-state communication no longer achieves the Holevo bound because number states are fragile under channel noise. A coherent-state input satisfies the minimum output entropy condition, and thus the Holevo quantity (\ref{eq:Holevo}) of coherent-state communication saturates the upper bound given by Eq. (\ref{eq:losscapacity}). However, the Holevo bound cannot be achieved with conventional Gaussian measurements such as homodyne and heterodyne measurements \cite{bib:PhysRevLett.92.027902,bib:PhysRevA.89.042309}. To find a practically achievable capacity, we need to understand the gap between the Holevo bound and the capacity of single-channel communication with feasible resources.

In this paper, we study single-channel communication capacity employing Gaussian operations only. Although the single-channel Gaussian communication does not attain the ultimate capacity saturating the Holevo bound, it has a practical importance because single-mode Gaussian operations are readily achievable in laboratory. Recently, Takeoka and Guha investigated Gaussian communication under a loss channel, which employs a coherent state as an input state \cite{bib:PhysRevA.89.042309}. They obtained an optimal strategy with restriction to coherent state inputs and Gaussian receivers. However, if we employ other Gaussian states as input, a better strategy might exist beating the coherent-state communication. It is already known that under an ideal channel, squeezed-state communication attains higher capacity, $ C^\textrm{sq} = \log_2(1+2\bar{n}) $ than the coherent-state communication, $ C^\textrm{coh} = \log_2(1+\bar{n}) $ \cite{bib:RevModPhys.66.481,bib:RevModPhys.58.1001}.

We first investigate the capacities of two well-known single-channel Gaussian communications, coherent-state and squeezed-state schemes, extending them to phase-insensitive Gaussian channels leading to loss and amplification, respectively. We note that some studies previously addressed single-channel Gaussian communications employing other than coherent state input under a lossless channel \cite{bib:RevModPhys.66.481,bib:RevModPhys.58.1001}, where noise effects are not taken into consideration. We find that, under a loss channel, there exists a critical value of photon number $n_c$ below which the squeezed-state communication beats the coherent-state communication. As the energy $\bar{n}$ increases, the coherent-state scheme can beat the squeezed-state scheme in a broad parameter region \cite{bib:PhysRevA.50.3295}. Furthermore, we consider a more general scenario, where we employ an arbitrary Gaussian measurement as well as an arbitrary Gaussian input state. We find that the maximum capacity is achieved either by the coherent-state communication or by the squeezed-state communication. We finally investigate how the gap in the capacity between a single-channel Gaussian communication and the Holevo bound appears.

\section{\label{sec:fundamental}Fundamentals of single-channel Gaussian communication}

\subsection{Gaussian phase-insensitive channels}
To begin with, we briefly introduce the description of bosonic systems. For more details, we refer to a review paper, e.g. \cite{bib:RevModPhys.84.621}. A bosonic system can be described by quadrature field operators $ \hat{\boldsymbol{\xi}} = \left( \hat{x}, \hat{p} \right)^T $, which satisfy the canonical commutation relation $ \left[ \hat{x}, \hat{p} \right] = i $. In particular, a Gaussian state is fully described by the first-order moments (averages) $ \bar{\boldsymbol{\xi}} \equiv \langle \hat{\boldsymbol{\xi}} \rangle $ and the second-order moments (variances) that can be compactly represented by a covariance matrix (CM) $\boldsymbol{\gamma}$, with its elements 
\begin{equation}
\gamma_{ij} = \frac{1}{2} \left\langle \hat{\xi}_i \hat{\xi}_j + \hat{\xi}_j \hat{\xi}_i \right\rangle - \bar{\xi}_i \bar{\xi}_j.
\end{equation}
The Wigner function of a Gaussian state can be written in terms of its first and second moments as
\begin{equation}
W(x,p) = \frac{1}{2\pi\det\boldsymbol{\gamma}} \exp\left[ -\frac{1}{2} \left(\boldsymbol{\xi}-\bar{\boldsymbol{\xi}}\right)^T \boldsymbol{\gamma}^{-1} \left(\boldsymbol{\xi}-\bar{\boldsymbol{\xi}}\right) \right] .
\end{equation}

\begin{figure}
\centering \includegraphics[width=0.8\columnwidth]{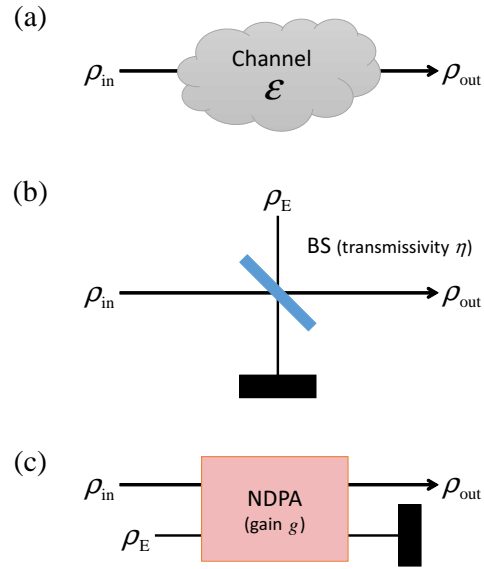}
\caption{\label{fig:channel} Representing (a) a general channel $\mathcal{E}$, (b) a loss channel described by a beam splitter (BS) interaction, and (c) an amplification channel described by a nondegenerate parametric amplifier (NDPA).}
\end{figure}

A deterministic quantum channel can generally be described by a positive, trace-preserving, map $\mathcal{E}$ that transforms a quantum state as $ \rho_\textrm{in} \to \rho_\textrm{out} = \mathcal{E}(\rho_\textrm{in}) $, as depicted in Fig. \ref{fig:channel}(a).
A loss channel can be represented by a beam-splitting interaction with a thermal reservoir field, as shown in \ref{fig:channel}(b), with the field operators transformed as
\begin{eqnarray}
\hat{x} & \to & \sqrt{\eta}\hat{x} + \sqrt{1-\eta}\hat{x}_E , \nonumber \\
\hat{p} & \to & \sqrt{\eta}\hat{p} + \sqrt{1-\eta}\hat{p}_E.
\end{eqnarray}
Here $ \eta$ $( 0 \leqslant \eta \leqslant 1 ) $ denotes the interaction strength with reservoir, with $\eta=0$ (1) corresponding to a complete loss (no loss), and $\hat{x}_E$ and $\hat{p}_E$ are reservoir operators with a thermal photon number $n_\textrm{th}$. In particular, when $ n_\textrm{th} = 0 $, the channel becomes a pure-loss channel. Let us consider an input state with the first moment $ \left( x, p \right)^T $ and CM
\begin{equation}
\boldsymbol{\gamma}^\textrm{sq} = \frac{1}{2} \left( \begin{array}{cc} e^{-2r} & 0 \\ 0 & e^{2r} \end{array} \right) ,
\end{equation}
which corresponds to a displaced squeezed state. Under a loss channel, the input state is transformed into an output state with the first moment $ \left( \sqrt{\eta}x, \sqrt{\eta}p \right)^T $ and CM
\begin{equation}
\boldsymbol{\gamma}^\textrm{sq}_\textrm{loss} = \frac{1}{2} \left( \begin{array}{cc} \eta e^{-2r} + N_\eta & 0 \\ 0 & \eta e^{2r} + N_\eta \end{array} \right) ,
\end{equation}
where $ N_\eta \equiv (1-\eta)(1+2n_\textrm{th}) $.

On the other hand, an amplification channel can be represented by a two-mode squeezing operation with a thermal reservoir, as shown in \ref{fig:channel}(c), which transforms field operators as
\begin{eqnarray} \label{eq:quadamp}
\hat{x} & \to & \sqrt{g}\hat{x} + \sqrt{g-1}\hat{x}_E , \nonumber \\
\hat{p} & \to & \sqrt{g}\hat{p} - \sqrt{g-1}\hat{p}_E ,
\end{eqnarray}
where $ g \geqslant 1$ denotes the intensity gain. A displaced squeezed state is transformed under an amplification channel into an output state with first moment $ \left( \sqrt{g}x, \sqrt{g}p \right)^T $ and CM
\begin{equation}
\boldsymbol{\gamma}^\textrm{sq}_\textrm{amp} = \frac{1}{2} \left( \begin{array}{cc} g e^{-2r} + N_g & 0 \\ 0 & g e^{2r} + N_g \end{array} \right) ,
\end{equation}
where $ N_g \equiv (g-1)(1+2n_\textrm{th}) $. Note that a most general phase-insensitive Gaussian channel can be represented by a concatenation of loss and amplification channels \cite{bib:NatPhoton.8.796,bib:NatCommun.5.3826}.

\subsection{Coherent-state communication}

In the coherent-state scheme, Alice prepares a vacuum state with zero means and CM
\begin{equation}
\boldsymbol{\gamma}^\textrm{coh} = \frac{1}{2} \left( \begin{array}{cc} 1 & 0 \\ 0 & 1 \end{array} \right) ,
\end{equation}
and encodes two variables $ \{ \alpha_x, \alpha_p \} $ by displacing her mode with an amplitude $ \frac{1}{\sqrt{2}} (\alpha_x+i\alpha_p)$ in phase space. The probability distribution for encoded $ \{ \alpha_x, \alpha_p \} $ is a Gaussian distribution given by
\begin{equation} \label{eq:symmetric}
P( \alpha_x, \alpha_p ) = \frac{1}{2\pi\sigma^2} \exp \left( -\frac{\alpha_x^2+\alpha_p^2}{2\sigma^2} \right) .
\end{equation}
The energy constraint on the channel input reads $ \bar{n} = \frac{1}{2} \left\langle \alpha_x^2+\alpha_p^2 \right\rangle = \sigma^2 $. Alice sends her mode to Bob, who subsequently performs heterodyne measurement. That is, Bob combines the received state with a vacuum state using a 50/50 beam splitter and measures $ \{ \beta_x, \beta_p \} $ on each output of the beam splitter, respectively. When Alice sends her mode via a loss channel, Bob's measurement result is centered at $ \left\{ \sqrt{\frac{\eta}{2}}\alpha_x, \sqrt{\frac{\eta}{2}}\alpha_p \right\} $ with variance
\begin{equation}
\Delta^2 \beta_x = \Delta^2 \beta_p = \frac{1}{2} \left( \frac{ \eta + N_\eta}{2} + \frac{1}{2} \right) ,
\end{equation}
where the two terms in the parentheses are the contributions from the received mode and the idler vacuum mode, respectively. Using the expression of mutual information for the case of Gaussian distribution in terms of noise $N$ and signal $S$ given by $ C = \frac{1}{2} \log_2 \left( 1+\frac{S}{N} \right) $ \cite{bib:InformationTheory}, we have the classical capacity of coherent-state communication in the loss channel as
\begin{eqnarray}
C_\textrm{loss}^\textrm{coh} & = & \frac{1}{2} \left[ \log_2 \left( 1+\frac{\frac{\eta}{2}\sigma^2}{\Delta^2\beta_x} \right) + \log_2 \left( 1+\frac{\frac{\eta}{2}\sigma^2}{\Delta^2\beta_p} \right) \right] \nonumber \\
& = & \log_2 \left( 1+\frac{2\eta\bar{n}}{1+\eta+N_\eta} \right) .
\end{eqnarray}
Note that $ \eta = 1 $ reproduces the capacity for a perfect channel $ C^\textrm{coh} = \log_2 (1+\bar{n}) $ and that capacity decreases as $\eta$ decreases or as $n_\textrm{th}$ increases. In a similar way, we find the classical capacity of coherent-state communication in the amplification channel as
\begin{equation}
C_\textrm{amp}^\textrm{coh} = \log_2 \left( 1+\frac{2g\bar{n}}{1+g+N_g} \right) .
\end{equation}

The above scheme takes into consideration a symmetric encoding on two orthogonal quadratures [Eq.~({\ref{eq:symmetric})], which results in a thermal state at the output, $ \sum_i p_i \mathcal{E}(\rho_i)= \rho_\textrm{th}$, thus achieving the Holevo bound. However, this does not automatically guarantee that the actually obtained mutual information is maximized as well under a Gaussian measurement (decoding) scheme. In the next section, we consider a general case where $ \left\langle \alpha_x^2 \right\rangle $ and $\left\langle \alpha_p^2 \right\rangle $ are not necessarily the same, and show that mutual information as well as the Holevo quantity is maximized by the symmetric encoding if both of the two quadratures are used for encoding on a vacuum state. Our result agrees with a recent work by Takeoka and Guha \cite{bib:PhysRevA.89.042309}, who showed that for coherent-state inputs, the symmetric two-quadrature encoding and heterodyne measurement is optimal in the large-photon-number regime $ \eta\bar{n} \geqslant \frac{2(1+n_\textrm{th})}{1+2n_\textrm{th}} $, while single-quadrature encoding and homodyne measurement is optimal in the small-photon-number regime $ \eta\bar{n} \leqslant \frac{2(1+n_\textrm{th})}{1+2n_\textrm{th}} $. Furthermore, we show that for the single-quadrature encoding with $\left\langle \alpha_p^2 \right\rangle=0$, a squeezed-state input attains a larger mutual information than a coherent-state input.

\subsection{Squeezed-state communication}

In the squeezed-state scheme, Alice prepares a squeezed state with zero means and CM $\boldsymbol{\gamma}^\textrm{sq}$ and encodes a single variable $\alpha_x$ only. Without loss of generality, we assume an $x$-squeezed state, i.e., $r>0$ in Eq. (7), and a displacement should then be performed along the $x$ quadrature having a smaller variance. We also assume that the probability distribution for $\alpha_x$ is Gaussian,
\begin{equation}
P(\alpha_x) = \frac{1}{\sqrt{2\pi}\sigma_x} \exp\left( -\frac{\alpha_x^2}{2\sigma_x^2} \right) .
\end{equation}
In this case, the photon number of input state is determined by two parameters, i.e., $ \bar{n} = n_0 + n_s $ where $ n_0 = \sinh^2 r $ is the photon number of the initial squeezed state and $ n_s = \frac{1}{2}\sigma_x^2 $ the photon number used for encoding. Bob reads $\beta_x$ by homodyne detection on the received state. In the case of loss channel, the measurement outcome is centered at $\sqrt{\eta}\alpha_x$ with its variance given by
\begin{equation}
\Delta^2 \beta_x = \left( \boldsymbol{\gamma}_\textrm{loss}^\textrm{sq} \right)_{11} = \frac{1}{2} \left( \eta e^{-2r} + N_\eta \right) .
\end{equation}
Then the capacity turns out to be
\begin{eqnarray}
C & = & \frac{1}{2} \log_2 \left( 1+\frac{\eta\sigma_x^2}{\Delta^2\beta_x} \right) \nonumber \\
& = & \frac{1}{2} \log_2 \left[ 1+\frac{4\eta(\bar{n}-\sinh^2 r)}{\eta e^{-2r} + N_\eta} \right] .
\end{eqnarray}
We find the optimal squeezing maximizing the above capacity under the energy constraint $\bar{n}$ by solving $ \partial C / \partial (e^{2r}) = 0 $, which leads to the solution
\begin{equation} \label{eq:sqopt}
\exp(2r) = \frac{-\eta+\sqrt{4\eta N_\eta\bar{n} + \left(N_\eta+\eta\right)^2}}{N_\eta}.
\end{equation}
The optimal squeezing gives the maximum capacity of squeezed-state communication in the loss channel as
\begin{eqnarray}
C_\textrm{loss}^\textrm{sq} & = & \frac{1}{2} \log_2\left[ \frac{ \left( -\eta+\sqrt{4\eta N_\eta\bar{n} + \left(N_\eta+\eta\right)^2} \right)^2 }{N_\eta^2} \right] \nonumber \\
& = & \log_2\left[ \frac{-\eta+\sqrt{4\eta N_\eta\bar{n} + \left(N_\eta+\eta\right)^2}}{N_\eta} \right] .
\end{eqnarray}
Similarly, the maximum capacity of squeezed-state communication in amplification channel is given by
\begin{equation}
C_\textrm{amp}^\textrm{sq} = \log_2\left[ \frac{-g+\sqrt{4g N_g\bar{n} + \left(N_g+g\right)^2}}{N_g} \right] .
\end{equation}

In general, the optimal squeezing (\ref{eq:sqopt}) does not result in a thermal state at the output. The ensemble of output states $ \sum_i p_i \mathcal{E}(\rho_i)$ becomes a thermal state only when the channel is perfect. As the channel becomes noisy, that is, as $\eta$ decreases ($g$ increases) or $n_\textrm{th}$ increases, the optimal squeezing decreases as well as the capacity decreases.

\subsection{Comparison between practical capacity and\\ Holevo quantity}

We now compare the practical capacities of two Gaussian communication protocols (coherent-state and squeezed-state schemes) together with the Holevo quantity given by Eq. (\ref{eq:Holevo}) under two Gaussian channels, loss and amplification, respectively (Fig. \ref{fig:capacity}).
\begin{figure*}
\centering \includegraphics[width=0.65\textwidth]{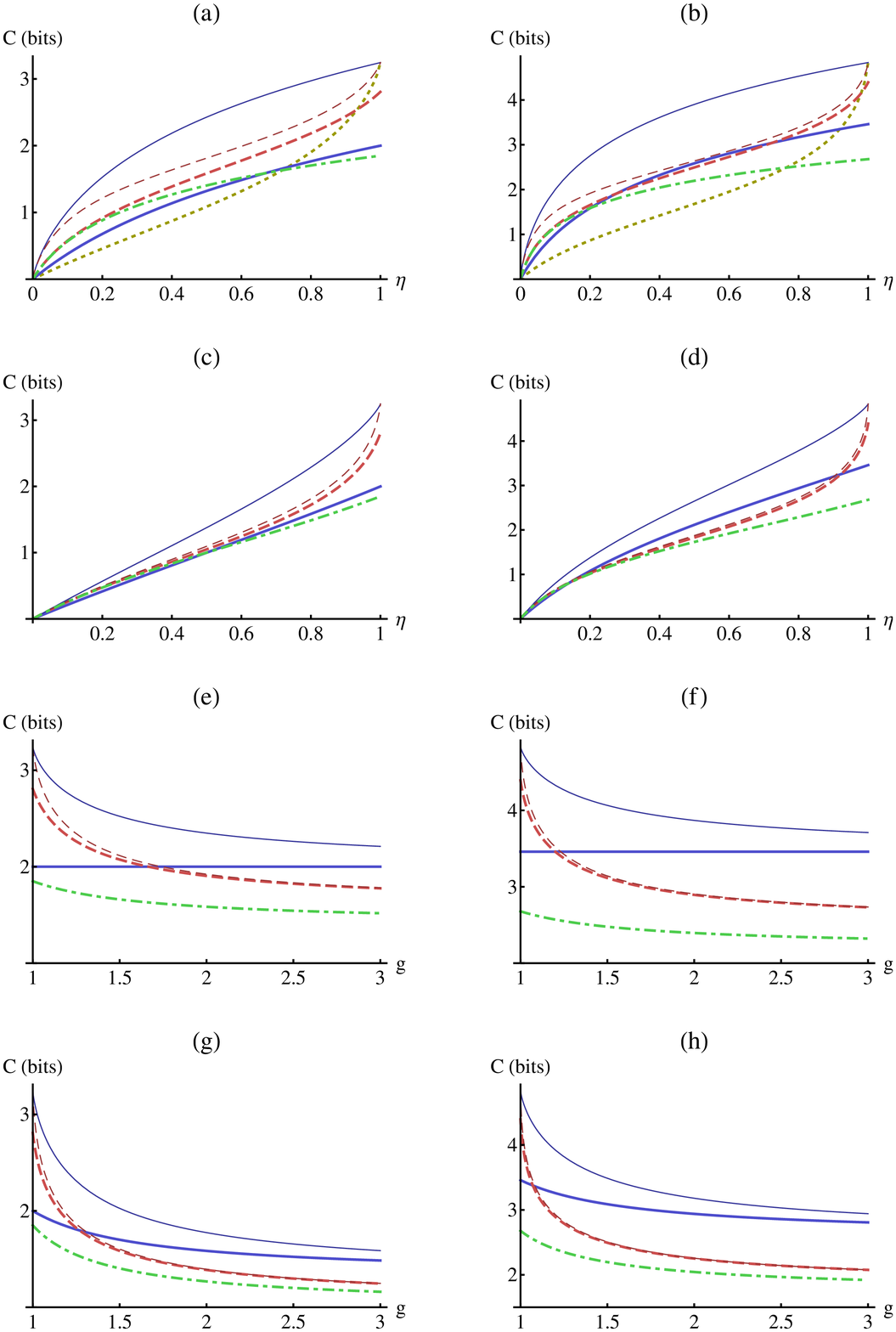}
\caption{\label{fig:capacity}Capacity for (a,b) pure-loss channel ($n_\textrm{th}=0$), (c,d) thermal-loss channel ($n_\textrm{th}=1$), (e,f) quantum-limited amplification channel ($n_\textrm{th}=0$), and (g,h) amplification channel with added noise ($n_\textrm{th}=1$). Energy constraint is given by $\bar{n}=3$ for left panels (a,c,e,f), and $\bar{n}=10$ for right panels (b,d,f,h), respectively. The practical capacities of coherent-state communication [Eqs. (14) and (15)] and those of squeezed-state communication [Eqs. (20) and (21)] are represented by a thick solid blue curve and a thick dashed red curve, respectively. Holevo quantities [Eq.(1)] for the coherent-state communication and squeezed-state communication are represented by thin solid blue curve and thin dashed red curve, respectively. We plot the capacity of another Gaussian communication protocol, i.e., the single-quadrature encoding on a coherent state (dot-dashed green curve). In (a) and (b), we also plot a numerically calculated capacity of number-state communication (dotted yellow curve). The uppermost curve (thin blue solid) represents the ultimate channel capacity whereas an optimal practical scheme appears right below it for each channel.}
\end{figure*}
We also consider the capacities of two other schemes. One is the number-state communication introduced in Sec. \ref{sec:introduction} (yellow dotted curves in Fig. \ref{fig:capacity}(a) and (b)). The other protocol is the single-quadrature encoding on a coherent state, which is also a widely studied Gaussian communication protocol \cite{bib:PhysRevA.89.042309,bib:RevModPhys.58.1001}. Obviously the capacity of the latter is always less than that of squeezed-state communication, which is the optimal strategy for the single-quadrature encoding. Fig. \ref{fig:capacity} (a)-(d) show that for a coherent-state input, the single-quadrature encoding is better than the two-quadrature encoding when $\eta\bar{n}$ is small. More precisely, as also identified in \cite{bib:PhysRevA.89.042309}, the range is given by $ \eta\bar{n} \leqslant \frac{2(1+n_\textrm{th})}{1+2n_\textrm{th}}$.

Let us first look into the Holevo quantities in Eq. (1). We find that the Holevo quantity of coherent-state communication (thin blue solid curves) is always greater than that of squeezed-state communication (thin red dashed curves), reminding us that a coherent state input leads to the Holevo bound (ultimate channel capacity) \cite{bib:NatPhoton.8.796,bib:NatCommun.5.3826}. The Holevo bound can be achieved only by the number-state communication (dotted yellow curves in Fig. \ref{fig:capacity}(a,b)) under a perfect channel with $ \eta = 1 $. As the channel becomes noisy, the actual capacity of number-state scheme decreases rapidly, whereas a Gaussian communication maintains a moderate level of capacity. This is because number states are perfectly distinguishable under a perfect channel, but highly fragile against the Gaussian channel noise.

The practical capacity of single-channel Gaussian communications (thick curves) is less than the Holevo quantity (thin curves) under each protocol. It becomes asymptotically close to the Holevo quantity in two cases. One is under the amplification channel in the limit of $ g \to \infty $, which is an unrealistic situation. The other is the squeezed-state communication with a sufficient noise. [Cf) thin and thick red dashed curves] Although the actual capacity of squeezed-state communication becomes nearly the same as the corresponding Holevo quantity with sufficient noise such that $ n_\textrm{th} > 0 $ and $ \eta \ll 1 $ (or $ g \gg 1 $), it is still less than the ultimate Holevo bound (thin solid curves) of each channel. Fig. 2 identifies an apparent gap between the practical capacities of two protocols and the ultimate Holevo bound. In the next section, we prove that these two protocols are optimal among general single-channel Gaussian communications.


Before moving on, let us remark on some features of each capacity. In a loss channel, signal intensity decreases to $0$ as $\eta$ goes to zero, while noise remains nonzero due to the channel noise or vacuum fluctuation even in a pure-loss channel. Thus the capacity for the loss channel goes to $0$ as $\eta$ goes to zero. On the other hand, the capacity for the amplification channel is constant or decreases monotonically to a finite value as $g$ goes to $\infty$ because both signal intensity and noise increase linearly with $g$. Especially, the capacity of coherent-state communication in a quantum-limited amplification channel is constant against $g$ (Fig. 2 (e) and (f)), that is, signal and noise increase at the same rate. Note that quantum amplification process always involves noise from environment as depicted in Eq. (\ref{eq:quadamp}). 
On the other hand, one expects an increase of capacity with amplification under a {\it classical} communication scheme, where signal can grow larger than noise. Such a classical behavior may emerge under certain conditions, for instance, when the heterodyne measurement at Bob's station introduces a larger variance $\sigma_m=\frac{w}{2}$ ($w>1$) than the ideal case $\sigma_m=\frac{1}{2}$, e.g. due to the coarse-grained measurement \cite{bib:PhysRevA.89.042102}. The capacity in Eq. (15) then reads 
$C_\textrm{amp}^\textrm{coh} = \log_2 \left( 1+\frac{2g\bar{n}}{w+g+N_g} \right)$, which monotonically increases with gain $g$ if $w>1+2n_\textrm{th}$, i.e., if the noise added through measurement is larger than the noise due to amplification.


Comparing the capacity of coherent-state communication and that of squeezed-state communication, although the former always maximizes the Holevo quantity, it does not necessarily maximize the practical capacity under Gaussian measurements (decoding). 
When $\bar{n}$ is rather small in a loss channel (Fig. \ref{fig:capacity}(a) and (c)), the squeezed-state communication always beats the coherent-state communication regardless of $\eta$. On the other hand, when $\bar{n}$ is large (right panels in Fig. \ref{fig:capacity}), the two schemes show crossings twice as $\eta$ varies from 0 to 1, which shows that the coherent-state scheme is better in an intermediate range of $\eta$. For further details, we plot the region where the coherent-state communication beats the squeezed-state communication in Fig. \ref{fig:region}.
\begin{figure}[t]
\centering \includegraphics[width=0.95\columnwidth]{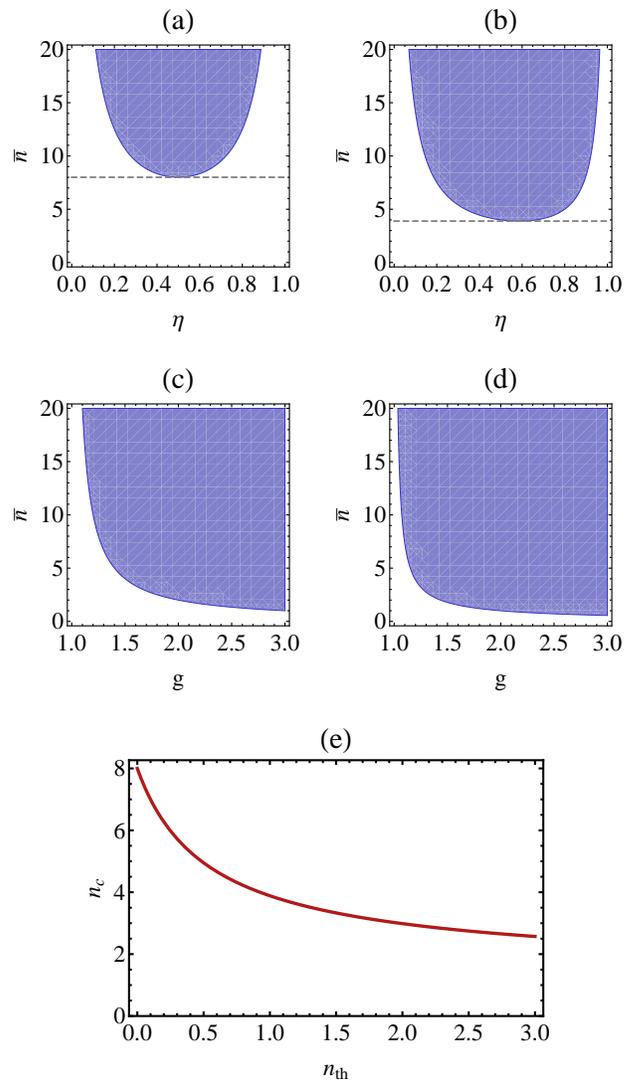}
\caption{\label{fig:region} Parameter region where the coherent-state communication beats the squeezed-state communication under (a) pure-loss channel, (b) thermal-loss channel with $ n_\textrm{th} = 1 $, (c) quantum-limited amplification channel, and (d) amplification with added noise $ n_\textrm{th} = 1 $. The horizontal dashed line represents the critical photon number $ \bar{n} = n_c $, below which the squeezed-state scheme always beats the coherent-state scheme. In (e), we plot the behavior of $n_c$ against $n_\textrm{th}$.}
\end{figure}
For a loss channel, a critical value of photon number, given by
\begin{equation}
n_c = \frac{4+2n_\textrm{th}+4\sqrt{1+n_\textrm{th}}}{1+2n_\textrm{th}} ,
\end{equation}
exists so that the squeezed-state communication always beats the coherent-state communication when $ \bar{n} < n_c $. For example, under a pure-loss channel with $n_\textrm{th}=0$, the coherent-state scheme can beat the squeezed-state scheme only for $ \bar{n} >8 $. As shown in Fig. \ref{fig:region} (e), $n_c$ decreases monotonically with $n_\textrm{th}$. As $\bar{n}$ increases above $n_c$, the coherent-state communication manifests a larger capacity in a wide range of $\eta$. For both of loss and amplification channels, the region becomes broader as $\bar{n}$ increases or as $n_\textrm{th}$ increases. This is also related to the fact that a coherent state suffers minimum disturbance from channel noise. With a large $\bar{n}$, the squeezed-state communication requires a large amount of squeezing [Eq.(19)], which also makes a squeezed state more disturbed by channel noise. Moreover, a squeezed state becomes more fragile as the channel has a larger $n_\textrm{th}$.

\section{\label{sec:general}Capacity of general single-channel Gaussian communication}

\subsection{General Gaussian communication protocol}

In this section, we investigate a generalized single-channel Gaussian communication, in which Alice employs a squeezed state input and Bob measures the outcome by projection onto a squeezed state. Here, we describe the case of loss channel only, but this method can also be straightforwardly extended to the amplification channel. In this generalized protocol, Alice prepares a squeezed state with squeezing strength $r$ and encodes two variables $ \{ \alpha_x, \alpha_p \} $ with the corresponding probability distribution given by
\begin{equation}
P( \alpha_x, \alpha_p ) = \frac{1}{2\pi\sigma_x \sigma_p} \exp \left( -\frac{\alpha_x^2}{2\sigma_x^2} -\frac{\alpha_p^2}{2\sigma_p^2} \right),
\end{equation}
which includes, as a special case, the encoding of a single-variable $\alpha_x$, i.e., $ \sigma_p = 0 $. Without loss of generality, we assume the input state to be an $x$-squeezed state ($r\geqslant0$). Alice sends her state to Bob via a noisy Gaussian channel and Bob receives a state $\rho_\textrm{out}$. Then Bob reads the outcome $ \{ \beta_x, \beta_p \} $ by a measurement with elements $\mathcal{M}_\beta$, which reads
\begin{equation}
\mathcal{M}_\beta = \frac{1}{\pi} \hat{D}\left({\textstyle\frac{1}{\sqrt{2}}}\beta\right) \ket{s} \bra{s} \hat{D}^\dagger\left({\textstyle\frac{1}{\sqrt{2}}}\beta\right).
\end{equation}
It describes a projection onto a displaced squeezed state with squeezing parameter $s$ and displacement $ \frac{1}{\sqrt{2}}\beta $ where $ \beta = \beta_x + i\beta_p $. As special cases, $s=0$ and $s\to\infty$ correspond to heterodyne detection and homodyne detection, respectively. The squeezing parameter $s$ of the measurement state is not necessarily the same as the squeezing parameter $r$ of the input state. Later, we also identify a relation between $r$ and $s$ for an optimal communication.

The variance of measurement outcome is determined by both the internal fluctuation of the received state and added noise from measurement \cite{bib:RevModPhys.58.1001}, that is, $ \Delta^2 \beta_x = \langle \Delta^2 x \rangle_{\rho_\textrm{out}} + \langle \Delta^2 x \rangle_{\rho_m} $, where $\rho_m$ is the state in Eq. (24), and similarly for $\Delta^2\beta_p$. If we employ a mixed state input, we can always find a pure squeezed state that results in an output state with smaller variances $ \langle x^2 \rangle_{\rho_\textrm{out}} $ and $ \langle p^2 \rangle_{\rho_\textrm{out}} $. Therefore, it suffices to consider only a pure state input, and similarly, to consider only the projection onto a pure state for optimization purpose. Then, the variance of measurement outcome, under a loss-channel with a thermal noise, is written as
\begin{eqnarray}
\Delta^2 \beta_x & = & \frac{\eta e^{-2r} + N_\eta}{2} + \frac{e^{-2s}}{2} , \nonumber \\
\Delta^2 \beta_p & = & \frac{\eta e^{2r} + N_\eta}{2} + \frac{e^{2s}}{2} .
\end{eqnarray}
Using again the expression of capacity $ C = \frac{1}{2} \log_2 \left( 1+\frac{S}{N} \right) $ for a Gaussian communication, we obtain
\begin{eqnarray}
C & = & \frac{1}{2}\left[ \log_2\left( 1 + \frac{\eta\sigma_x^2}{\Delta^2 \beta_x} \right) + \log_2\left( 1 + \frac{\eta\sigma_p^2}{\Delta^2 \beta_p} \right) \right] \\
& = & \frac{1}{2}\left[ \log_2\left( 1 + \frac{2\eta\sigma_x^2}{N_\eta + \eta e^{-2r} + e^{-2s}} \right) \right. \nonumber \\
& & \quad \left. + \log_2\left( 1 + \frac{2\eta\sigma_p^2}{N_\eta + \eta e^{2r} + e^{2s}} \right) \right] \nonumber ,
\end{eqnarray}
with the energy constraint
\begin{equation} \label{eq:constraint}
\bar{n} = \frac{1}{2}(\sigma_x^2+\sigma_p^2) + \sinh^2 r .
\end{equation}
In the case when Alice encodes a single variable ($ \sigma_p = 0 $), Bob's best strategy is to read only the $x$ quadrature with a minimum disturbance, which corresponds to homodyne measurement ($ s \to \infty $). This strategy is exactly the same as the squeezed-state communication we have discussed in the last section, which leads to the capacity $C_\textrm{loss}^\textrm{sq}$ in Eq. (20).

\subsection{Optimization}

Now we are going to obtain a maximum capacity by the following steps. In a first step, we optimize $\sigma_x$ and $\sigma_p$ for given $r$ (squeezing for encoding) and $s$ (squeezing for decoding). When Alice encodes variables, she can adjust the degree of encoding on two quadratures within the photon number constraint (\ref{eq:constraint}). We find the optimal encoding as
\begin{eqnarray} \label{eq:encoding}
(\sigma_x^2)_\textrm{opt} & = & \bar{n} - \sinh^2 r + \frac{1}{2}\sinh(2r) + \frac{1}{2\eta}\sinh(2s) , \nonumber \\
(\sigma_p^2)_\textrm{opt} & = & \bar{n} - \sinh^2 r - \frac{1}{2}\sinh(2r) - \frac{1}{2\eta}\sinh(2s) ,
\end{eqnarray}
and the corresponding capacity as
\begin{equation}
C' = \log_2\left[ \frac{\eta+2\eta\bar{n}+N_\eta+\cosh(2s)}{\sqrt{(N_\eta+\eta e^{-2r}+e^{-2s})(N_\eta+\eta e^{2r}+e^{2s})}}\right] .
\end{equation}
Eq. (\ref{eq:encoding}) indicates that we need to encode more information on the $x$-quadrature that can be measured more accurately. This optimal scheme involves the two-quadrature encoding only when $ (\sigma_p^2)_\textrm{opt} > 0 $, or,
\begin{equation} \label{eq:condition}
\bar{n} - \sinh^2 r > \left| \frac{1}{2}\sinh(2r) + \frac{1}{2\eta}\sinh(2s) \right| .
\end{equation}
If the above inequality does not hold, the optimal strategy becomes the single-quadrature encoding, that is, the squeezed-state communication. We have already found the optimal solution for the case of single-quadrature encoding in Sec. II C, i.e., Eqs. (18)-(20) together with homodyne detection. Therefore, we now focus on the optimization of two-quadrature encoding strategy that is relevant only to the case satisfying the energy condition in Eq. (30).

In a second step, we find the optimal value of $r$ in terms of $s$. Using $ z+\frac{1}{z} \geqslant 2 $, it is easy to find that $C'$ is maximized when
\begin{equation} \label{eq:ropt}
\exp(2r_\textrm{opt}) = \sqrt{\frac{N_\eta+e^{2s}}{N_\eta+e^{-2s}}} ,
\end{equation}
and the corresponding optimal capacity becomes
\begin{equation}
C'' = \log_2\left( \frac{\eta+2\eta\bar{n}+N_\eta+\cosh(2s)}{\eta+\sqrt{1+N_\eta^2+2N_\eta\cosh(2s)}} \right) .
\end{equation}
Eq. (\ref{eq:ropt}) provides a recipe for an optimal preparation of the initial state for a given receiver. For instance, when Bob performs heterodyne measurement ($s=0$), it is optimal to prepare a coherent state ($r_\textrm{opt}=0$). Combining Eqs. (\ref{eq:encoding}) and (\ref{eq:ropt}), we similarly find that the optimal strategy for a coherent state input reads $ \sigma_x = \sigma_p $ and $s=0$, that is, the coherent-state communication employing a symmetric encoding and the heterodyne detection, as addressed in the last section, is optimal for a coherent state input. 

If the channel is perfect ($\eta=1$), Eq. (31) gives $r_\textrm{opt}=s$. That is, when Bob performs measurement via projection onto a squeezed state, Alice must prepare a squeezed state with the same squeezing parameter. However, in a noisy channel, $r_\textrm{opt}$ does not exactly coincide with $s$, but increases monotonically with $s$.

The last step for a full optimization is to find the optimal value of $s$. Within the range given in Eq. (\ref{eq:condition}), there exist at most three extremal points: ({\romannumeral 1}) $ \cosh(2s) = \frac{1}{N_\eta}{\scriptstyle \left[ -1 + \eta \left( \eta+2\eta\bar{n}+N_\eta - \sqrt{-1+\eta^2+N_\eta(2\eta+4\eta\bar{n}+N_\eta)} \right) \right]} $, ({\romannumeral 2}) $s=0$, and the ({\romannumeral 3}) boundary of (\ref{eq:condition}) combined with (\ref{eq:ropt}). We may ignore the point ({\romannumeral 1}) which gives only a local minimum that may exist or not. The point ({\romannumeral 2}) $s=0$ gives a local maximum if the point ({\romannumeral 1}) exists and it corresponds to the coherent-state communication. If we have a maximum at the point ({\romannumeral 3}), it actually corresponds to the case of single-quadrature encoding and thus we can achieve a larger capacity with the optimal squeezed-state communication. Therefore, we conclude that the maximum capacity of single-channel Gaussian communication is achieved either by the coherent-state communication or by the squeezed-state communication, that is,
\begin{equation}
C_\textrm{loss}^\textrm{G} = \max \left\{ C_\textrm{loss}^\textrm{coh} , C_\textrm{loss}^\textrm{sq} \right\} .
\end{equation}
Which of the two protocols gives a larger capacity depends on the input energy $\bar{n}$ as well as the characteristics of the channel, as we have already shown in Fig. \ref{fig:capacity}. Because either of these two Gaussian protocols does not achieve the Holevo bound, we also conclude that no single-channel Gaussian communication is sufficient to achieve the ultimate Holevo bound. 
We finally show the photon number efficiency, which quantifies the number of bits per photon, in Fig. \ref{fig:efficiency}.
\begin{figure}[t]
\centering \includegraphics[width=0.7\columnwidth]{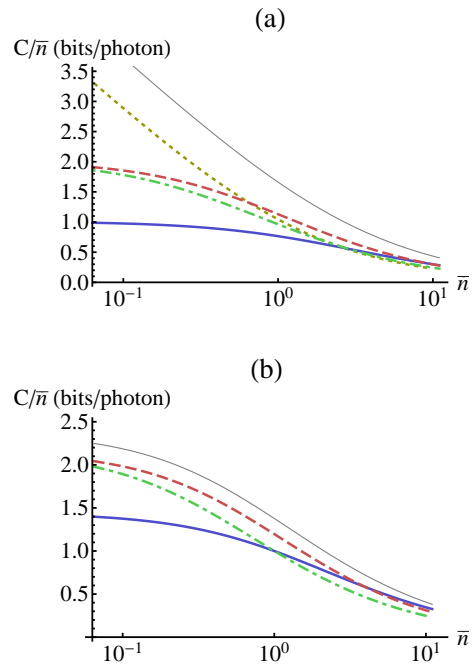}
\caption{\label{fig:efficiency}Plot of photon information efficiency for (a) pure-loss channel with $\eta=0.7,n_\textrm{th}=0$, and (b) quantum-limited amplification channel with $g=1.5,n_\textrm{th}=0$. Each curve represents the Holevo bound (uppermost thin curve), the capacity of coherent-state communication (blue solid), the capacity of squeezed-state communication (red dashed), the capacity of number-state communication (yellow dotted), and the capacity of single-quadrature encoding on a coherent state (green dot-dashed).}
\end{figure}
We find an apparent gap between the Holevo bound and the capacity of single-channel Gaussian communication. For a loss channel, the gap becomes large as $\bar{n}$ goes to $0$, while the number-state communication works quite well. For an amplification channel, the gap is rather small, but still nonzero.

\section{\label{sec:conclusion}Conclusion}

In this paper, we have studied the capacities of Gaussian communications under a single noisy Gaussian channel. We have proved that for a given channel with an energy constraint, the optimal protocol is either the coherent-state communication or the squeezed-state communication among generalized Gaussian schemes. In a small-photon-number regime ($ \bar{n} < n_c $) under a loss channel, the squeezed-state communication is always optimal regardless of loss rate, $1-\eta$. However, as $\bar{n}$ or $n_\textrm{th}$ increases, the coherent-state communication may attain a greater capacity in a broad parameter region. The superiority of squeezed-state scheme to coherent-state scheme also emerges under an amplification channel in a small photon-number regime.

On the other hand, we have also investigated the gap in capacity between an optimal Gaussian communication (readily accessible in laboratory) and the information-theoretic Holevo bound in a broad region of parameters.
A future work should of course be directed to narrow this gap by a practically feasible scheme. Although the recent proof on the minimum output conjecture gives the ultimate channel capacity achievable, it is not yet known how close to this ultimate capacity one can experimentally reach. The HSW theorem suggested an asymptotic method of achieving the Holevo bound by using an infinite number of channels, which should be more elaborated to the case of using a finite-number of channels. As our work clearly shows the limitation of using only Gaussian measurements for a Gaussian-state communication under a single noisy channel, it must be further extended to non-Gaussian operations and measurements, which will be studied in future.

Furthermore, our study should also be extended to the case of using multiple channels together with a collective measurement.
Recently, it was shown that, with coherent-state input and Gaussian receiver only, a separable receiver is optimal, that is, collective measurement does not make any improvement \cite{bib:PhysRevA.89.042309}. On the other hand, for communication of discrete random variables using a finite number of coherent states, there was a theoretical proposal on how to construct joint-detection receivers that achieves the Holevo bound \cite{bib:PhysRevLett.106.240502}. It has also been reported that separable measurements with feedforward improves the sensitivity in discriminating codewords so as to approach the Holevo bound. There has been experimental demonstration of joint-detection receivers which discriminate codewords encoded as sequences of coherent states, with error rates below the standard quantum limit achievable with heterodyne measurement \cite{bib:NatPhoton.6.374,bib:NatPhoton.9.48}.
Building upon the results of our current work, we plan to investigate Gaussian communications involving collective operations. It is still yet to be known that any collective Gaussian measurement yields an improvement for continuous-variable communication with general Gaussian input states other than coherent states. If it turns out that any Gaussian communication is optimal with a separable receiver, our work already reveals a clear gap between the ultimate Holevo bound and the capacity achievable within the Gaussian regime.

\section*{acknowledgement}
This work is supported by the NPRP grant 4-554-1-084 from Qatar National Research Fund.

\end{document}